\documentclass[twocolumn,showpacs,preprintnumbers,amsmath,amssymb]{revtex4-1}
\usepackage[colorlinks,linkcolor=blue,anchorcolor=blue,citecolor=blue]{hyperref}
\usepackage{graphicx,epsfig}% Include figure files
\usepackage{dcolumn}% Align table columns on decimal point
\usepackage{bm}% bold math
\usepackage{fixfoot}
\usepackage{etoolbox}
\usepackage{kbordermatrix}
\begin{document}

\title{Analytical solutions for quantum walks on 1D chain with different shift operators}
\author{Xin-Ping Xu$^{1,2}$}
%\email{xuxp@ihep.ac.cn \\
% Tel:+82-70-76014610}
\author{Xiao-Kun Zhang$^1$}
\author{Yusuke Ide$^{3}$}

\author{Norio Konno$^{4}$}

\affiliation{%
$^1$School of Physical Science and Technology, Soochow University, Suzhou 215006, China\\
$^2$Department of Physics and Astronomy, Seoul National University, Seoul 151-747, Korea \\
$^3$ Department of Information Systems Creation, Faculty of Engineering, Kanagawa University, Yokohama, Kanagawa, 221-8686, Japan\\
$^4$ Department of Applied Mathematics, Faculty of Engineering, Yokohama National University, Hodogaya, Yokohama 240-8501, Japan
}%
%\date{\today}
\begin{abstract}
In this paper, we study the discrete-time quantum walks on 1D Chain with the moving and swapping shift operators. We derive analytical solutions for the eigenvalues and eigenstates of the evolution operator $\hat{U}$ using the Chebyshev polynomial technique, and calculate the long-time averaged probabilities for the two different shift operators respectively. It is found that the probability distributions for the moving and swapping shift operators display completely different characteristics. For the moving shift operator, the probability distribution exhibits high symmetry where the probabilities at mirror positions are equal. The probabilities are inversely proportional to the system size $N$ and approach to zero as $N\rightarrow \infty$. On the contrary, for the swapping shift operator, the probability distribution is not symmetric, the probability distribution approaches to a power-law stationary distribution as $N\rightarrow \infty$ under certain coin parameter condition. We show that such power-law stationary distribution is determined by the eigenstates of the eigenvalues $\pm1$ and calculate the intrinsic probability for different starting positions. Our findings suggest that the eigenstates corresponding to eigenvalues $\pm1$ play an important role for the swapping shift operator.
\end{abstract}
 \maketitle
\section{Introduction}
Quantum walks have become a popular research topic in the past few years~\cite{rn1,rn2,rn3,rn4,rn5}. The continuous interest in quantum walk can be attributed to its broad applications to many distinct fields, such as polymer physics, solid state physics, biological physics, and quantum computation~\cite{rn6,rn7}. In the literature~\cite{rn1,rn2,rn3}, there are two types of quantum walks: continuous-time and discrete-time quantum walks. The main difference of the two types of quantum walks is that discrete-time quantum walk (DTQW) requires an extra coin Hilbert space in which the coin operator acts, while continuous-time quantum walks (CTQWs) do not need this extra Hilbert space. Aside from this, these two quantum walks (QWs) are similar to their classical counterparts. Discrete-time quantum walks evolve by the application of a unitary evolution operator at discrete time intervals, and continuous-time quantum walks evolve under a (usually time-independent) Hamiltonian in Schr\"{o}dinger picture. Due to the different dimensional Hilbert space, CTQWs can't be regarded as the limit of DTQWs as the time step goes to zero and there is no simple relation connecting the two quantum walk models~\cite{add1,add2}.

Here, we focus on the discrete-time quantum walk (DTQWs). Previous work have studied DTQWs on simple regular structures, for instance, the line~\cite{rn8}, cycle~\cite{rn9,xuxp} and hypercube~\cite{rn10}. For DTQWs on the line, the problem could be simplified using the Fourier transform technique. In Ref.~\cite{rn9}, Bednarska {\it et al.} have studied the DTQWs using a Hadamard coin on the cycle. Various methods such as Schr\"{o}dinger picture~\cite{rn11}, combinatorial approach~\cite{rn12,rn13,rn14}, generating function~\cite{rn15}, scattering theory~\cite{rn16}, {\it etc}, are employed to treat the problem.
In Refs.~\cite{andris,bach}, the authors have studied the 1D DTQWs with one and two absorbing boundaries. They introduce absorbing boundary by implementing a measurement operator to the quantum state and derive an absorption probability for the boundaries using a Hadamard coin. On the contrary, in this paper we concentrate on the pure quantum evolution process without measurement and disturbance. We will study the DTQWs on the one-dimensional (1D) chain, which lacks one connection compared to the cycles. The left-most and right-most points are topological boundaries of the 1D chain and will impose some influences to the quantum dynamics. Although there are some studies of 1D DTQWs in the literature~\cite{ide}, DTQWs on 1D chain has received little attention due to the difficult analytical calculation. In this work, we will shed some light on this problem. We study DTQWs on 1D chain with two different shift operators: the moving and swapping shift operators. Both of the shift operators are used in the community, the moving shift operator acting on a state only move the position of the particle while the swapping shift operator changes both the position and direction of the particle's quantum state. The swapping operator can be applied to any graphs and widely used in the Szegedy's quantum walk~\cite{rn17}, which is a generalized discrete quantum walk defined on general graphs via the quantization of a stochastic matrix. We will derive analytical results for DTQWs on 1D chain with the moving and swapping shift operators, and compare the difference between them.

The rest of the paper is organized as follows. Section II introduces the model of DTQWs on general graphs. Section III gives the theoretical framework of DTQWs on 1D chain with moving and swapping shift operators, and determines the matrix form of the various shift operators and evolution operators in the Hilbert space. In Sec. IV, we show the analytical results for the moving shift operator. We determine the eigenvalues and eigenstates of the evolution operator and obtain the long-time averaged probability. In Sec. V, we show the analytical results for the swapping shift operator. We determine the eigenvalues and eigenstates of the evolution operator and calculate the long-time averaged probability. We also show that the eigenstates corresponding to the eigenvalues $\pm 1$ play an important role in the quantum dynamics. We compare the dynamical difference between the two shift operators. Conclusions and discussions are given in the last part, Sec. VI.

\section{Definition of discrete-time quantum walks}
%\subsection{Discrete-time quatum walks}
Discrete-time quantum walk was first introduced by Mayer and Aharonov {\it et al.} in Ref.~\cite{rn18,rn19}. DTQW takes place in a discrete position space, with a unitary evolution of coin toss and position shift in discrete time steps. Here, we review the definition of DTQWs on $d$-regular graph, which is a regular graph each vertex has exactly $d$ edges.

DTQW on $d$-regular graph happens on the coin Hilbert space ${\cal H}_c$ and position Hilbert space ${\cal H}_p$, the total Hilbert space is given by ${\cal H}={\cal H}_c\otimes {\cal H}_p$~\cite{rn1}. If the $d$-regular graph has $N$ vertices, the position and coin Hilbert space are denoted as ${\cal H}_p= \{|i\rangle: i=1,2,...,N\}$, ${\cal H}_c= \{|J\rangle: J=1,2,...,d\}$. The coin flip operator $\hat{C}$ and position shift operator $\hat{S}$ are applied to the
total state in ${\cal H}$ at each time step~\cite{rn1}. The coin flip operation $\hat{C}$ (acting on ${\cal H}_c$) is the quantum equivalent of randomly choosing which way the particle will move, then the position-shift operation $\hat{S}$ moves the particle according to the coin state, transferring the particle
into the new superposition state in position space. For every vertex, all the outgoing edges are labeled as $1,2,\ldots, d$. Let us call $e_i^J$ an edge $e=(i, i')$ which on $i$'s end is labelled by J. The conditional shift operation $\hat{S}$ moves the particle from $i$ to $i'$ if the edge $(i\rightarrow i')$ is labeled by $J$ on $i$'s side~\cite{rn1}:
\begin{equation} \label{eq1}
\hat{S}|i,J\rangle = \left\{
\begin{array}{ll}
|i',J \rangle, \text{ if }  e_i^J=(i,i'),  \\
0,  \text{ otherwise}.
\end{array}
\right.
\end{equation}
The evolution of the system at each step of the walk is governed by
the total operator,
\begin{equation}\label{eq2}
\hat{U}=\hat{S}(\hat{I}_p\otimes \hat{C}),
\end{equation}
where $\hat{I}$ is the identity operator in ${\cal H}_p$. Thus the
total state after $t$ steps is given by,
\begin{equation}\label{eq3}
|\psi(t)\rangle= \hat{U}^t |\psi(0)\rangle,
\end{equation}
where $|\psi(0)\rangle$ is the initial state. Finally, we obtain the probability distribution,
\begin{equation}\label{eq4}
P(i,t)=\sum_{J=1}^d |\langle i,J|\psi(t)\rangle|^2= \sum_{J=1}^d
|\langle i,J|\hat{U}^t|\psi(0)\rangle|^2.
\end{equation}
Suppose the eigenvalue equation of $\hat{U}$ is $\hat{U}|\Psi_{i',J'}\rangle=u_{i',J'}|\Psi_{i',J'}\rangle$ ($i'\in[1,N], J'\in[1,d]$), where $u_{i',J'}$ and $|\Psi_{i',J'}\rangle$ are the eigenvalues and orthonormalized eigenstates of the evolution operator $\hat{U}$. Then Eq.~(\ref{eq4}) can be written as,
\begin{equation}\label{eq5}
\begin{aligned}
P(i,t)&=\sum_{J=1}^d \Big |\sum_{J'=1}^d \sum_{i'=1}^N u_{i',J'}^t \langle i,J|\Psi_{i',J'}\rangle \langle \Psi_{i',J'}|\psi(0)\rangle \Big |^2\\
&=\sum_{J=1}^d \sum_{i',J'}\sum_{i'',J''}u_{i',J'}^tu_{i'',J''}^{*t}  \langle i,J|\Psi_{i',J'}\rangle \langle \Psi_{i',J'}|\psi(0)\rangle  \\
&\ \ \ \ \ \ \ \ \ \ \ \ \ \ \ \ \ \ \ \ \ \ \ \ \ \langle\psi(0)|\Psi_{i'',J''}\rangle \langle \Psi_{i'',J''}|i,J\rangle
.
\end{aligned}
\end{equation}
Noting that $\hat{U}$ is a unitary operator, {\it i.e.}, $\hat{U}\hat{U}^{\dag}=\hat{I}$ (which leads to $|u|=1$), the long time averages of $P(i,t)$ can be written as,
\begin{equation}\label{eq6}
\begin{aligned}
\chi(i)&=\lim_{T\rightarrow\infty}\frac{1}{T}\sum_{t=0}^TP(i,t) \\
&=\sum_{J=1}^d \sum_{i',J'}\sum_{i'',J''} \langle i,J|\Psi_{i',J'}\rangle \langle \Psi_{i',J'}|\psi(0)\rangle  \\
&\langle\psi(0)|\Psi_{i'',J''}\rangle \langle \Psi_{i'',J''}|i,J\rangle \lim_{T\rightarrow\infty}\frac{1}{T}\sum_{t=0}^T (u_{i',J'}u_{i'',J''}^*)^t \\
&=\sum_{J=1}^d \sum_{i',J'}\sum_{i'',J''}\delta(u_{i',J'}-u_{i'',J''}) \langle i,J|\Psi_{i',J'}\rangle \langle \Psi_{i',J'}|\psi(0)\rangle  \\
&\langle\psi(0)|\Psi_{i'',J''}\rangle \langle \Psi_{i'',J''}|i,J\rangle,
\end{aligned}
\end{equation}
where $\delta(u_{i',J'}-u_{i'',J''})$ takes value 1 if $u_{i',J'}=u_{i'',J''}$ and equals to 0 otherwise. In the above equation, we can see that the limit distribution $\chi(i)$ depends on the eigenvalues and eigenstates of the evolution operator $\hat{U}$. If all the eigenvalues $\{ u_{i',J'}|i'\in[1,N], J'\in[1,d] \}$ are different ({\it i.e.}, all the eigenvalues are not degenerated), the above Equation can be simplified as,
\begin{equation}\label{eq7}
\begin{aligned}
\chi(i)=\sum_{J=1}^d \sum_{i',J'} |\langle i,J|\Psi_{i',J'}\rangle|^2\cdot |\langle \Psi_{i',J'}|\psi(0)\rangle|^2  .
\end{aligned}
\end{equation}
In order to calculate the analytical expressions for $P(i,t)$ and $\chi(i)$, all the eigenvalues $u_{i',J'}$ and eigenstates $|\Psi_{i',J'}\rangle$
of the evolution operator $\hat{U}$ are required. In the following we will use Eq.~(\ref{eq7}) to derive the limit probability distribution of DTQWs on 1D chain, and put emphasis on the calculation of the eigenvalues and eigenstates of $\hat{U}$.

\section{DTQWs on 1D chain with the moving and swapping shift operators}
In this section, we will continue the calculation for DTQWs on 1D chain. For a 1D chain system composed of $N$ nodes, which are labeled as $\{i:i=1,2,...,N\}$,
each node $1<i<N$ is connected two nearest neighbors ($i-1$ and $i+1$) except the leftmost node $1$ and rightmost node $N$. The Hilbert space has $2N-2$ base vectors, which are denoted as $|2,L\rangle, |3,L\rangle, \cdots, |N,L\rangle, |1,R\rangle, |2,R\rangle, \cdots, |N-1,R\rangle$. We will derive the matrix form for the coin operator and shift operator in the Hilbert space, and obtain the matrix representation for the evolution operator. We also determine the eigenvalues and eigenstates of $\hat{U}$ based on the matrix representation.
\subsection{Coin operator $\hat{C}$ and initial state}
The probability distribution is related to the coin operator $\hat{C}$ and initial state $|\psi(0)\rangle$. In the scientific community, the coin flip operator $\hat{C}$ can be of various forms, for instance, the most commonly used Grover coin and the Discrete Fourier Transform (DFT) coin~\cite{rn1,rn2}. It has been shown that different choices of coin flip operator $\hat{C}$ and initial state $|\psi(0)\rangle$ may have different dynamics. Here, we will specify the form for the coin operator $\hat{C}$ and initial state $|\psi(0)\rangle$.

For the coin operator, we use the coin type of DTQWs in Szegedy's scheme~\cite{szegedy}, which can be generalized to high dimensional case. If the particle is located at the two end points ($i=1,N$), the particle will move to the neighboring node in the next step (do not need coin). If the particle is located at the center nodes $1<i<N$, we use an arbitrary coin state  $|\phi\rangle=\cos\frac{1}{2}\omega|L\rangle+e^{i\varphi}\sin\frac{1}{2}\omega|R\rangle$ to generate the coin operator,
\begin{equation}\label{eq8}
\begin{aligned}
\hat{C}=&2|\phi\rangle\langle\phi|-I=\cos\omega|L\rangle\langle L|+ e^{i\varphi}\sin\omega|L\rangle\langle R| \\
&+e^{-i\varphi}\sin\omega|R\rangle\langle L|  -\cos\omega|R\rangle\langle R|
\end{aligned}
\end{equation}
When $\varphi=0$, the above coin operator becomes
\begin{equation}\label{eq9}
\begin{aligned}
\hat{C}=&  \begin{bmatrix}
       \cos\omega|L\rangle\langle L| &\ \ \ \sin\omega |L\rangle\langle R| \\
       \sin\omega|R\rangle\langle L| & -\cos\omega|R\rangle\langle R|
       \end{bmatrix} \\
\equiv & \begin{bmatrix}
       a|L\rangle\langle L| &\ \ \ b|L\rangle\langle R| \\
       b|R\rangle\langle L| &-a|R\rangle\langle R|
       \end{bmatrix}, \ \ \ \ a\equiv\cos\omega, \  b\equiv\sin\omega.
\end{aligned}
\end{equation}

The direct product of $\hat{I}_p$ and $\hat{C}$ can be written as,
\begin{equation}\label{eq10}
\begin{aligned}
\hat{C}_p=\hat{I}_p\otimes\hat{C}=&|1\rangle\langle1|\otimes|R\rangle\langle R|+|N\rangle\langle N|\otimes|L\rangle\langle L| \\
&+\sum_{i=2}^{{\scriptscriptstyle N}-1}|i\rangle\langle i|\otimes\hat{C}.
\end{aligned}
\end{equation}
In the Hilbert space, the above coin operator $\hat{C}_p$ can be represented as the following matrix,
\begin{equation}\label{eq11}
\hat{C}_p=
\kbordermatrix{\mbox{}&2&\cdots&{\scriptscriptstyle N-1}&{\scriptscriptstyle N}&\vrule &1&2&\cdots&{\scriptscriptstyle N-1}\\
\cline{2-10}
2                        &a         &       &\mathbf{0} &0      &\vrule &0      &b         &       & \mathbf{0}     \\
\vdots                   &          &\ddots &           &\vdots &\vrule &\vdots &          &\ddots &         \\
{\scriptscriptstyle N-1} &\mathbf{0}&       &a          &0      &\vrule &0      &\mathbf{0}&       &b      \\
{\scriptscriptstyle N}   &0         &\cdots &0          &1      &\vrule &0      &0         &\cdots &0      \\
\hline
1                        &0         &\cdots &0          &0      &\vrule &1      &0         &\cdots &0      \\
2                        &b         &       &\mathbf{0} &0      &\vrule &0      &-a        &       &\mathbf{0}      \\
\vdots                   &          &\ddots &           &\vdots &\vrule &\vdots &          &\ddots &      \\
{\scriptscriptstyle N-1} &\mathbf{0}&       &b          &0      &\vrule &0      &\mathbf{0}&       &-a
},
\end{equation}
where the four squares correspond to the orthonormalized basis set $|L\rangle\langle L|$, $|L\rangle\langle R|$, $|R\rangle\langle L|$, $|R\rangle\langle R|$ respectively. In this paper, we will use this coin operator to do analytical calculations.

For the initial state $|\psi(0)\rangle=|i_0\rangle\otimes|C_0\rangle$, the initial position $i_0$ can be at the two end points and the center nodes, and the corresponding initial coin state $|C_0\rangle$ can be single state ($|L\rangle \text{ or } |R\rangle$) and superposed state $(\cos\omega_0|L\rangle+e^{i\varphi_0}\sin\omega_0|R\rangle)$. Consequently, the initial state  $|\psi(0)\rangle$ can be summarized as follows,
\begin{equation} \footnotesize  \label{eq12}
|\psi(0)\rangle=
\begin{cases}
|1\rangle\otimes|R\rangle=|1,R\rangle, & \text{if }i_0=1 \\
|N\rangle\otimes|L\rangle=|N,L\rangle, & \text{if }i_0=N \\
|i_0\rangle\otimes(\cos\omega_0|L\rangle+e^{i\varphi_0}\sin\omega_0|R\rangle), &i_0\in(1,N)
\end{cases}
\end{equation}
Here, for the sake of simplicity, we set $\omega_0=\pi/4$ and choose appropriate $\varphi_0$ value to make the product $|\langle \Psi_{i,J}|\psi(0)\rangle|^2$ equals to $|\langle\Psi_{i_0,L}|i_0,L\rangle|^2\cdot|\langle i_0,L|\psi (0)\rangle|^2+|\langle\Psi_{i_0,R}|i_0,R\rangle|^2\cdot|\langle i_0,R|\psi (0)\rangle|^2$ ($i_0\in (1,N)$).

\subsection{The moving and swapping shift operators}
In this section, we define the moving and swapping shift operators. The moving shift operator $\hat{S}^m$ moves the particle to the neighboring position and keep the direction $J$ unchanged. This means $\hat{S}^m|i,L\rangle=|i-1,L\rangle$ ($i\in [3,N]$), $\hat{S}^m|i,R\rangle=|i+1,R\rangle$ ($i\in [1,N-2]$). For the two boundary positions, we use the elastic reflection condition: $\hat{S}^m|2,L\rangle=|1,R\rangle$, $\hat{S}^m|N-1,R\rangle=|N,L\rangle$. Thus, the moving shift operator acting on $|i,J\rangle$ is summarized as follows,
\begin{small}
\begin{equation}\label{eq13}
\hat{S}^m|i,J\rangle=
\begin{cases}
|i+1,J\rangle,  & \text{if } J=R \text{ and } 1\leqslant i\leqslant N-2 \\
|i-1,J\rangle,  & \text{if } J=L \text{ and } 3\leqslant i\leqslant N   \\
|N,L\rangle,  & \text{if } J=R \text{ and } i=N-1         \\
|1,R\rangle,  & \text{if } J=L \text{ and } i=2
\end{cases}
\end{equation}
\end{small}
Hence, the element of $\hat{S}^m$ in the Hilbert space is,
\begin{equation}\footnotesize \label{eq14}
\begin{aligned}
\langle i,J|\hat{S}^m|i',J'\rangle=
\begin{cases}
\delta_{i+1,i'},  \text{if } J=L, J'=L \\
\delta_{i,i'+1},  \text{if } J=R, J'=R \\
1,  \text{if } |i,J\rangle=|{\scriptscriptstyle N,L}\rangle, |i',J'\rangle=|{\scriptscriptstyle N-1,R}\rangle \\
1,  \text{if } |i,J\rangle=|1,R\rangle, |i',J'\rangle=|2,L\rangle \\
0,  \text{Otherwise.}
\end{cases}
\end{aligned}
\end{equation}
For brevity, we represent the moving shift operator $\hat{S}^m$ in the Hilbert space using the following matrix,
\begin{equation} \label{eq15}
\hat{S}^m=
\kbordermatrix{\mbox{}&2&\cdots&{\scriptscriptstyle N-1}&{\scriptscriptstyle N}&\vrule &1&2&\cdots&{\scriptscriptstyle N-1}\\
\cline{2-10}
2                        &0         &1      &           &\mathbf{0}&\vrule &0    &0         &0       &0         \\
\vdots                   &\vdots    &       &\ddots     &       &\vrule &0       &0         &0       &0         \\
{\scriptscriptstyle N-1} &0         &\mathbf{0}&        &1      &\vrule &0      &0         &0         &0        \\
{\scriptscriptstyle N}   &0         &\cdots &0          &0      &\vrule &0      &0         &0         &1      \\
\hline
1                        &1         &0      &0          &0      &\vrule &0      &0         &\cdots    &0      \\
2                        &0         &0      &0          &0      &\vrule &1      &          &\mathbf{0}&0      \\
\vdots                   &0         &0      &0          &0      &\vrule &       &\ddots    &       &\vdots     \\
{\scriptscriptstyle N-1} &0         &0      &0          &0      &\vrule &\mathbf{0}&       &1      &0
}
\end{equation}

The swapping shift operator swaps the particle's state, moving the particle to the neighboring position and changing the direction.
The swapping shift operator does not need boundary condition and can be applied to arbitrary graphs. Specifically, the swapping shift
operator $\hat{S}^s$ acting on $|i,J\rangle$ is summarized as,
\begin{equation}\label{eq16}
\hat{S}^s|i,J\rangle=
\begin{cases}
|i+1,L\rangle,  & \text{if } J=R \\
|i-1,R\rangle,  & \text{if } J=L
\end{cases}
\end{equation}
Thus, the element of $\hat{S}^s$ in the Hilbert space is,
\begin{equation}\label{eq17}
\begin{aligned}
\langle i,J|\hat{S}^s|i',J'\rangle=
\begin{cases}
\delta_{i,i'+1},  & \text{if } J=L, J'=R \\
\delta_{i+1,i'},  & \text{if } J=R, J'=L \\
0, & \text{Otherwise.}
\end{cases}
\end{aligned}
\end{equation}
In the meantime, the matrix form for the swapping shift operator $\hat{S}^s$ is,
\begin{equation}\label{eq18}
\hat{S}^s=
\kbordermatrix{\mbox{}&2&\cdots&{\scriptscriptstyle N-1}&{\scriptscriptstyle N}&\vrule &1&2&\cdots&{\scriptscriptstyle N-1}\\
\cline{2-10}
2                        &          &       &           &       &\vrule &1      &          &       & \mathbf{0}     \\
\vdots                   &          &       &           &       &\vrule &       &\ddots    &       &         \\
{\scriptscriptstyle N-1} &          &       &\mathbf{0} &       &\vrule &       &          &\ddots &      \\
{\scriptscriptstyle N}   &          &       &           &       &\vrule &\mathbf{0}&       &       &1      \\
\hline
1                        &1         &       &           &\mathbf{0}&\vrule &      &          &       &       \\
2                        &          &\ddots &           &       &\vrule &       &\mathbf{0}&       &        \\
\vdots                   &          &       &\ddots     &       &\vrule &       &          &       &      \\
{\scriptscriptstyle N-1} &\mathbf{0}&       &           &1      &\vrule &       &          &       &
}
\end{equation}

\subsection{Evolution operators $\hat{U}$ for DTQWs with the moving and swapping shift operators}
In this section, we will use the matrix form of the coin operator $\hat{C}_p$ (Eq.~(\ref{eq11})) and shift operators (Eq.~(\ref{eq15}) and (\ref{eq18})) to calculate the evolution operator $\hat{U}$.

For the moving shift operator, the evolution operator $\hat{U}^m$ can be obtained by calculating the product of $\hat{S}^m$ and $\hat{C}_p$, which leads to,
\begin{footnotesize}
\begin{equation}\label{eq19}
\hat{U}^m=\hat{S}^m\hat{C}_p=
\kbordermatrix{\mbox{}&2&3&\cdots&{\scriptscriptstyle N-1}&{\scriptscriptstyle N}&\vrule &1&2&\cdots&{\scriptscriptstyle N-2}&{\scriptscriptstyle N-1}\\
\cline{2-12}
2                        &0         &a      &           &          &   &\vrule &0    &0          &b       &      &        \\
3                        &          &0      &\ddots     &          &   &\vrule &     &0          &0       &\ddots&         \\
\vdots                   &          &       &0          &a         &   &\vrule &     &           &0       &0     &b         \\
{\scriptscriptstyle N-1} &          &       &           &0         &1  &\vrule &     &           &        &0     &0        \\
{\scriptscriptstyle N}   &          &       &           &b         &0  &\vrule &     &           &        &      &-a        \\
\hline
1                        &a         &       &           &          &   &\vrule &0    &b          &        &       &      \\
2                        &0         &0      &           &          &   &\vrule &1    &0          &        &       &      \\
\vdots                   &b         &0      &0          &          &   &\vrule &     &-a         &0       &       &     \\
{\scriptscriptstyle N-2} &          &\ddots &0          &0         &   &\vrule &     &           &\ddots  &0      &     \\
{\scriptscriptstyle N-1} &          &       &b          &0         &0  &\vrule &     &           &        &-a     &0
}
\end{equation}
\end{footnotesize}

Analogously, for the swapping shift operator, the evolution operator $\hat{U}^s$ is calculated to be,
\begin{footnotesize}
\begin{equation}\label{eq20}
\hat{U}^s=\hat{S}^s\hat{C}_p=
\kbordermatrix{\mbox{}&2&\cdots&{\scriptscriptstyle N-1}&{\scriptscriptstyle N}&\vrule &1&2&\cdots&{\scriptscriptstyle N-1}\\
\cline{2-10}
2                        &0         &       &           &       &\vrule &1      &          &       & \mathbf{0}     \\
\vdots                   &b         &0      &           &       &\vrule &       &-a        &       &         \\
{\scriptscriptstyle N-1} &          &\ddots &0          &       &\vrule &       &          &\ddots &      \\
{\scriptscriptstyle N}   &          &       &b          &0      &\vrule &\mathbf{0}&       &       &-a      \\
\hline
1                        &a         &       &           &\mathbf{0}&\vrule &0   &b         &       &       \\
2                        &          &\ddots &           &       &\vrule &       &0         &\ddots &     \\
\vdots                   &          &       &a          &       &\vrule &       &          &0      &b      \\
{\scriptscriptstyle N-1} &\mathbf{0}&       &           &1      &\vrule &       &          &       &0
}
\end{equation}
\end{footnotesize}

It is worth mentioning that we have obtained the matrix form of the evolution operator $\hat{U}^m$ for the moving shift operator, as well as the evolution operator $\hat{U}^s$ for the swapping shift operator. As we can see, the two evolution operators are different and may lead to complete different quantum dynamics. This is the key issue we are trying to reveal in this paper.

\section{Results for the moving shift operator}
In this section, we determine the eigenvalues and eigenstates of the evolution operator $\hat{U}^m$ for DTQWs using the moving shift operator. We also use the eigenvalues and eigenstates to calculate the probability distribution in Eq.~(\ref{eq7}).
\subsection{Eigenvalues and eigenstates of $\hat{U}^m$}
We start our analysis on the eigen equation of the evolution operator $\hat{U}^m$ for DTQWs with the moving shift operator. Suppose the eigen equation of $\hat{U}^m$ is $\hat{U}^m|\Psi^m\rangle=u|\Psi^m\rangle$ ($\hat{U}^m|\Psi_{i,J}^m\rangle=u_{i,J}^m|\Psi_{i,J}^m\rangle$), the eigenstates $|\Psi^m\rangle$ can be expressed as
\begin{equation}\small    \label{eq21}
\begin{aligned}
|\Psi^m\rangle=&\sum_{i,J}\alpha_{i,J}^m|i,J\rangle\\
=&\alpha_{2,L}^m|1,L\rangle+\alpha_{3,L}^m|2,L\rangle+\cdots +\alpha_{N,L}^m|N,L\rangle \\
&+\alpha_{1,R}^m|1,R\rangle+\alpha_{2,R}^m|2,R\rangle+\cdots +\alpha_{N-1,R}^m|N-1,R\rangle
\end{aligned}
\end{equation}

Noting that the matrix form of $\hat{U}^m$ in Eq.~(\ref{eq19}), the eigen equation $\hat{U}^m|\Psi^m\rangle=u|\Psi^m\rangle$ can be decomposed into the following $2N-2$ linear equations,
\begin{align}
a\alpha_{i+1,L}^m+b\alpha_{i+1,R}^m&=u\alpha_{i,L}^m,  \ \ \ 2\leqslant i\leqslant N-2 \label{eq22} \\
\alpha_{N,L}^m&=u\alpha_{N-1,L}^m, \    \label{eq23}  \\
b\alpha_{N-1,L}^m-a\alpha_{N-1,R}^m&=u\alpha_{N,L}^m, \label{eq24} \\
a\alpha_{2,L}^m+b\alpha_{2,R}^m&=u\alpha_{1,R}^m, \label{eq25} \\
\alpha_{1,R}^m&=u\alpha_{2,R}^m, \    \label{eq26} \\
b\alpha_{i,L}^m-a\alpha_{i,R}^m&=u\alpha_{i+1,R}^m,  \ \ \ 2\leqslant i\leqslant N-2 \label{eq27}
\end{align}
Utilizing Eq.~(\ref{eq22}) to eliminate $\alpha_{i,R}^m$ and $\alpha_{i+1,R}^m$, Eq.~(\ref{eq27}) becomes $\alpha_{i+1,L}^m-\alpha_{i-1,L}^m=\frac{u^2-1}{au}\alpha_{i,L}^m$. This is similar to the recursive relation of the variant Chebyshev
polynomials (see Appendix B). Noting the recursive relations and the mapping relationship $\frac{u^2-1}{au}\equiv 2y$ in the definition
of the variant Chebyshev polynomials, the variables $\alpha_{i,L}^m$ ($i\in [2,N-1]$) can be expressed as a function of $\alpha_{3,L}^m$ and $\alpha_{2,L}^m$,
\begin{equation}\label{eq28}
\alpha_{i,L}^m=V_{i-3}(y)\alpha_{3,L}^m+V_{i-4}(y)\alpha_{2,L}^m, \ \ i\in [2,N-1]
\end{equation}
where $V_{n}(y)=i^nU_{n}(x)$ ($y=ix$) is the variant Chebyshev polynomials (See Appendix B). Utilizing Eq.~(\ref{eq22}) to eliminate $\alpha_{N-1,R}^m$ in Eqs.~(\ref{eq23}), (\ref{eq24}), we get the relationship of $\alpha_{N-1,L}^m$ and $\alpha_{N-2,L}^m$,
\begin{equation}\label{eq29}
(1-bu^2)\alpha_{N-1,L}^m-au\alpha_{N-2,L}^m=0.
\end{equation}
Combining Eqs.~(\ref{eq28}) and (\ref{eq29}), we have,
\begin{equation}\label{eq30}
\begin{aligned}
&[(1-bu^2)V_{N-4}(y)-auV_{N-5}(y)]\alpha_{3,L}^m\\
&-[auV_{N-6}(y)-(1-bu^2)V_{N-5}(y)]\alpha_{2,L}^m=0
\end{aligned}
\end{equation}
In order to get another equation for $\alpha_{3,L}^m$ and $\alpha_{2,L}^m$, we combine Eqs.~(\ref{eq25}) and (\ref{eq26}),
\begin{equation}\label{eq31}
\alpha_{2,R}^m=\frac{a\alpha_{2,L}^m}{u^2-b},\ \ \  \alpha_{1,R}^m=\frac{au\alpha_{2,L}^m}{u^2-b}.
\end{equation}
Setting $i=2$ in Eq.~(\ref{eq27}), we obtain $b\alpha_{2,L}^m-a\alpha_{2,R}^m=u\alpha_{3,R}^m$. Substituting $\alpha_{2,R}^m$ in Eq.~(\ref{eq31}) and $\alpha_{3,R}^m$ in Eq.~(\ref{eq22}) ($i=2$) into this equation, we get another equation for $\alpha_{3,L}^m$ and $\alpha_{2,L}^m$,
\begin{equation}\label{eq32}
au(u^2-b)\alpha_{3,L}^m
-[(u^2-b)(u^2-b^2)+a^2b]\alpha_{2,L}^m=0
\end{equation}
Thus we have got two equations for $\alpha_{3,L}^m$ and $\alpha_{2,L}^m$, Eq.~(\ref{eq30}) and Eq.~(\ref{eq32}). The two equations should have nonzero solutions, hence the determinant of the four coefficients equals to 0. In the Appendix C, we show that the substraction of the coefficient products can be simplified to be a much simple form in Eq.~(\ref{c10}), thus we have obtained determinant equation for the eigenvalues,
\begin{equation}\label{eq33}
U_{{\scriptscriptstyle N}-1}(x)+\frac{b-1}{b+1}U_{{\scriptscriptstyle N}-3}(x)=0,
\end{equation}
where $U_{n}(x)$ are Chebyshev polynomials of the second kind.

There is no exact analytical solution for Eq.~(\ref{eq33}). However, when the system size $N$ is large, the solution for Eq.~(\ref{eq33}) is close to the equation $U_{{\scriptscriptstyle N}-1}(x)=0$, which leads to $N-1$ roots $x_k=\cos\theta_k, \theta_k=k\pi/N, k=1,2,\cdots,N-1$. Using the mapping relation $u=ay\pm\sqrt{1+a^2y^2}=axi\pm\sqrt{1-a^2x^2}$, the $2N-2$ eigenvalues of $\hat{U}^m$ are given by,
\begin{equation}\label{eq34} \small
u_{\pm k}=ax_k i\pm\sqrt{1-a^2x_k^2}, x_k=\cos \frac{k\pi}{N}, k\in [1,N-1].
\end{equation}

Now we analyze the eigenstates $|\Psi^m\rangle$. For this purpose, we denote all the components $\alpha_{i,J}^m$ in terms of $\alpha_{2,L}^m$. First, according to Eq.~(\ref{eq32}), we can express $\alpha_{3,L}^m$ as a function of $\alpha_{2,L}^m$,
\begin{equation}\label{eq35}
\begin{aligned}
\alpha_{3,L}^m=&\frac{(u^2-b)(u^2-b^2)+a^2b}{au(u^2-b)}\alpha_{2,L}^m  \\
=&\Big[\frac{u^2-b^2}{au}+\frac{ab}{u(u^2-b)}\Big]\alpha_{2,L}^m \\
=&\Big[\frac{(u^2-1)+(1-b^2)}{au}+\frac{ab}{u(u^2-b)}\Big]\alpha_{2,L}^m \\
=&\Big[\frac{2iaux+a^2}{au}+\frac{ab}{u(u^2-b)}\Big]\alpha_{2,L}^m \\
=&\Big(2ix+\frac{au}{u^2-b}\Big)\alpha_{2,L}^m,
\end{aligned}
\end{equation}
where $a^2+b^2=1$ and mapping relation $\frac{u^2-1}{au}\equiv 2y=2ix$ are used. Substituting Eq.~(\ref{eq35}) into Eq.~(\ref{eq28}), we have,
\begin{equation}\label{eq36}
\begin{aligned}
\alpha_{j,L}^m=&\Big[i^{j-3}U_{j-3}(x)(2ix+\frac{au}{u^2-b})+i^{j-4}U_{j-4}(x)\Big]\alpha_{2,L}^m \\
=&i^j\Big[\frac{aui}{u^2-b}U_{j-3}(x)-U_{j-2}(x)\Big]\alpha_{2,L}^m    \\
=&-i^j\frac{auiU_{j-1}(x)+(1-b)U_{j-2}(x)}{u^2-b}\alpha_{2,L}^m, \ \ \ \ j\in [2,N-1]
\end{aligned}
\end{equation}
where identity (\ref{a7}) and $\frac{u^2-1}{au}\equiv 2y=2ix$ are applied in the above calculation. According to Eq.~(\ref{eq23}), we have
\begin{equation}\label{eq37}
\begin{aligned}
\alpha_{N,L}^m=&u\alpha_{N-1,L}^m\\
=&ui^{N-1}\Big[\frac{aui}{u^2-b}U_{N-4}(x)-U_{N-3}(x)\Big]\alpha_{2,L}^m
\end{aligned}
\end{equation}
Likewise, substituting Eq.~(\ref{eq36}) into Eq.~(\ref{eq22}) and utilizing identity (\ref{a7}) and the mapping relation, we have
\begin{equation}\label{eq38}
\begin{aligned}
\alpha_{j,R}^m=&\frac{1}{b}\Big\{ i^{j-1}u [\frac{aui}{u^2-b}U_{j-4}(x)-U_{j-3}(x)]  \\
& -i^{j}a[\frac{aui}{u^2-b}U_{j-3}(x)-U_{j-2}(x)]  \Big\} \alpha_{2,L}^m \\
=&-i^j\frac{(1-b)uiU_{j-3}(x)+aU_{j-2}(x)}{u^2-b}\alpha_{2,L}^m \\
\end{aligned}
\end{equation}
The above equation holds for $j\in[2,N-1]$. According to Eq.~(\ref{eq26}), we have
\begin{equation}\label{eq39}
\alpha_{1,R}^m=u\alpha_{2,R}^m=\frac{au}{u^2-b}\alpha_{2,L}^m.
\end{equation}
We have obtained all the eigenstates as a function of $\alpha_{2,L}^m$. Combining $U_{N-1}(x)+U_{N-3}(x)=2xU_{N-2}(x)$ and Eq.~(\ref{eq33}) leads to
$U_{N-1}(x)=(1-b)xU_{N-2}(x)$, $U_{N-3}(x)=(1+b)xU_{N-2}(x)$. Noting that $U_{N-2}^2(x)-U_{N-1}(x)U_{N-3}(x)=1$, we arrive at the relation $(1-a^2x^2)U_{N-2}^2(x)=1$. Using these relations, one can prove the following symmetric properties for the eigenstates components,
\begin{align}
|\alpha_{1,R}^m|&=|\alpha_{2,R}^m|=|\alpha_{N-1,L}^m|=|\alpha_{N,L}^m| \label{eq40} \\
|\alpha_{i,L}^m|&=|\alpha_{i+1,R}^m| \ \ \ \ i\in[2,N-2] \label{eq41} \\
|\alpha_{i,L}^m|&=|\alpha_{N+1-i,R}^m| \ \ \ \ i\in[2,N]  \label{eq42}
\end{align}
$\alpha_{2,L}^m$ can be determined by the normalization condition $\sum_{i,J}|\alpha_{i,J}^m|^2=\sum_{i=2}^{N}|\alpha_{i,L}^m|^2+\sum_{i=1}^{N-1}|\alpha_{i,R}^m|^2=1$. After some algebraic calculus, we find an appropriate expression for $\alpha_{2,L}^m$,
\begin{equation}\label{eq43}
|\alpha_{2,L}^m|^2\approx \frac{1-b+4b(1+b)\cos^2\theta_k}{N[1+b^2+(b^2-1)\cos2\theta_k]}\sin^2\theta_k\sim\frac{1}{N}
\end{equation}

\subsection{Long-time averaged probability distribution}
Now, we consider the long-time averaged probability distribution. According to Eq.~(\ref{eq7}), the distribution can be written as,
\begin{equation}\label{eq44}
\begin{aligned}
\chi_{i,j}=& \sum_{|k|= 1}^{N-1}|\langle j, L|\Psi^m(x_k)\rangle|^2\cdot |\langle \Psi^m(x_k)|\psi_{i,0}\rangle|^2 \\
          & + \sum_{|k|= 1}^{N-1}|\langle j, R|\Psi^m(x_k)\rangle|^2\cdot |\langle \Psi^m(x_k)|\psi_{i,0}\rangle|^2.
\end{aligned}
\end{equation}
In the following, we will discuss $\chi_{i,j}$ according to the staring position and initial state $|\psi_{i,0}\rangle$ in Eq.~(\ref{eq12}). For the sake of simplicity, we first focus on the case $i=1$ (starting at the left most node). The initial state $|\psi_{i,0}\rangle=|1,R\rangle$ (see Eq.~(\ref{eq12})), the probability of finding the particle at the original site $j=i=1$ only depends on the second term of Eq.~(\ref{eq42}),
\begin{equation}\label{eq45}
\begin{aligned}
\chi_{1,1}=&\sum_{|k|= 1}^{N-1}|\langle 1, R|\Psi^m(x_k)\rangle|^4=2\sum_{k=1}^{N-1} |\alpha^m_{1,R}(x_k)|^4  \\
=&2\sum_{k=1}^{N-1} |\frac{au_k}{u^2_k-b}|^4\cdot|\alpha^m_{1,L}(x_k)|^4 \\
=&2\sum_{k=1}^{N-1} \frac{a^4}{(1-b)^2[1-b+4b(1+b)\cos^2\theta_k]^2} \\
& \cdot\frac{[1-b+4b(1+b)\cos^2\theta_k]^2\sin^4\theta_k}{N^2[1+b^2+(b^2-1)\cos2\theta_k]^2}\\
&\approx \frac{1}{N}\cdot\frac{2+b(b^2-3)}{4(b-1)^2}\sim\frac{1}{N}
\end{aligned}
\end{equation}
The probability of finding the particle at the right most site $j=N$ only depends on the first term of Eq.~(\ref{eq44}),
\begin{equation}\label{eq46}
\begin{aligned}
\chi_{1,N}=&\sum_{|k|= 1}^{N-1}|\langle N, L|\Psi^m(x_k)\rangle|^2\cdot |\langle \Psi^m(x_k)|1,R\rangle|^2 \\
=&2\sum_{k=1}^{N-1} |\alpha^m_{N,L}(x_k)|^2\cdot |\alpha^m_{1,R}(x_k)|^2 \\
=&2\sum_{k=1}^{N-1} |\alpha^m_{1,R}(x_k)|^4\equiv\chi_{1,1}
\end{aligned}
\end{equation}
Similarly, the probability of finding the particle at the middle sites $j\in[2,N-1]$ is related to both the two terms in Eq.~(\ref{eq42}),
\begin{equation}\label{eq47}
%\small
\begin{aligned}
\chi_{1,j}=&\sum_{|k|= 1}^{N-1}(|\alpha^m_{j,L}(x_k)|^2+|\alpha^m_{j,R}(x_k)|^2)\cdot |\alpha^m_{1,R}(x_k)|^2 \\
=&2\sum_{k= 1}^{N-1}(|\alpha^m_{j,L}(x_k)|^2+|\alpha^m_{j,R}(x_k)|^2)\cdot |\alpha^m_{1,R}(x_k)|^2
\end{aligned}
\end{equation}
Using the expression of $\alpha^m_{j,J}(x_k)$ in the above section, one can prove that $\chi_{1,j}$ is inversely proportion to the system size $N$, {\it i.e.}, $\chi_{1,j}\sim\frac{1}{N}$. Here, we do not show the details of the calculations.

If the starting position $i$ is not at the end points, the probabilities are also inversely proportion to the system size $N$. Therefore, for the moving shift operator, all the probabilities satisfy $\chi_{i,j}\sim 1/N$, which indicates that the probabilities approach to zero when the system size $N\rightarrow \infty$. This feature is quite different from the swapping shift operator, where the probabilities approach to a nonzero value when $N\rightarrow \infty$. Another interesting characteristic for the long-time averaged probability is that $\chi_{i,j}$ displays a symmetric relation: $\chi_{i,j}\equiv \chi_{i,N+1-j}$. This could be proved by the symmetric properties for the eigenstates components in Eqs.~(\ref{eq40})-(\ref{eq42}).

\section{Results for the swapping shift operator}
In this section, we determine the eigenvalues and eigenstates of the evolution operator $\hat{U}^s$ for DTQWs using the swapping shift operator. We use the eigenvalues and eigenstates to calculate the probability distribution in Eq.~(\ref{eq7}).
\subsection{Eigenvalues and eigenstates of $\hat{U}^s$}
We start our analysis on the eigen equation of the evolution operator $\hat{U}^s$ for DTQWs with the swapping shift operator. Suppose the eigen equation of $\hat{U}^s$ is $\hat{U}^s|\Psi^s\rangle=u|\Psi^s\rangle$ ($\hat{U}^s|\Psi_{i,J}^s\rangle=u_{i,J}^s|\Psi_{i,J}^s\rangle$), the eigenstates $|\Psi^s\rangle$ can be expressed as
\begin{equation}\small    \label{eq48}
\begin{aligned}
|\Psi^s\rangle=&\sum_{i,J}\alpha_{i,J}^s|i,J\rangle\\
=&\alpha_{2,L}^s|1,L\rangle+\alpha_{3,L}^s|2,L\rangle+\cdots +\alpha_{N,L}^s|N,L\rangle \\
&+\alpha_{1,R}^s|1,R\rangle+\alpha_{2,R}^s|2,R\rangle+\cdots +\alpha_{N-1,R}^s|N-1,R\rangle
\end{aligned}
\end{equation}

Noting that the matrix form of $\hat{U}^s$ in Eq.~(\ref{eq20}), the eigen equation $\hat{U}^s|\Psi^s\rangle=u|\Psi^s\rangle$ can be decomposed into the following $2N-2$ linear equations,
\begin{align}
\alpha_{1,R}^s&=u\alpha_{2,L}^s, \    \label{eq49}  \\
b\alpha_{i,L}^s-a\alpha_{i,R}^s&=u\alpha_{i+1,L}^s,  \ \ \ 2\leqslant i\leqslant N-1 \label{eq50} \\
a\alpha_{i,L}^s+b\alpha_{i,R}^s&=u\alpha_{i-1,R}^s,  \ \ \ 2\leqslant i\leqslant N-1 \label{eq51} \\
\alpha_{N,L}^s&=u\alpha_{N-1,R}^s. \    \label{eq52}
\end{align}
Utilizing Eq.~(\ref{eq50}) to eliminate $\alpha_{i,R}^s$ and $\alpha_{i-1,R}^s$, Eq.~(\ref{eq51}) becomes $\alpha_{i+1,L}^s+\alpha_{i-1,L}^s=\frac{u^2+1}{bu}\alpha_{i,L}^s$. This is similar to the recursive relation of the Chebyshev
polynomials (see Appendix A). Noting the recursive relations and the mapping relationship $\frac{u^2+1}{bu}\equiv 2x$ in the definition
of the Chebyshev polynomials of the second kind, the variables $\alpha_{i,L}^s$ ($i\in [2,N]$) can be expressed as a function of $\alpha_{3,L}^s$ and $\alpha_{2,L}^s$,
\begin{equation}\label{eq53}
\alpha_{i,L}^s=U_{i-3}(x)\alpha_{3,L}^s-U_{i-4}(x)\alpha_{2,L}^s, \ \ i\in [2,N]
\end{equation}
where $U_{n}(x)$ is the Chebyshev polynomials of the second kind (See Appendix A). Applying Eq.~(\ref{eq53}) to Eq.~(\ref{eq52}), we get the relationship of $\alpha_{3,L}^s$ and $\alpha_{2,L}^s$,
\begin{equation}\label{eq54}
\begin{aligned}
& [(a+u^2)U_{N-4}(x)-buU_{N-5}(x)]\alpha_{2,L}^s \\
& -[(a+u^2)U_{N-3}(x)-buU_{N-4}(x)]\alpha_{3,L}^s=0.
\end{aligned}
\end{equation}

According to Eq.~(\ref{eq51}), we have $a\alpha_{2,L}^s+b\alpha_{2,R}^s=u\alpha_{1,R}^s$. Utilizing Eq.~(\ref{eq50}) to eliminate $\alpha_{2,R}^s$ and noting $\alpha_{1,R}^s=u\alpha_{2,L}^s$ in Eq.~(\ref{eq49}), we get another equation for $\alpha_{3,L}^s$ and $\alpha_{2,L}^s$,
\begin{equation}\label{eq55}
(1-au^2)\alpha_{2,L}^s-bu\alpha_{3,L}^s=0
\end{equation}
Thus we have got two equations for $\alpha_{3,L}^s$ and $\alpha_{2,L}^s$, Eq.~(\ref{eq54}) and Eq.~(\ref{eq55}). The two equations should have nonzero solutions, the determinant of the four coefficients equals to 0. In the Appendix D, we show that the substraction of the coefficient products can be simplified to be a much simple form in Eq.~(\ref{a7}), thus we have obtained determinant equation for the eigenvalues,
\begin{equation}\label{eq56}
abu(u^2-1)U_{{\scriptscriptstyle N}-2}(x)=0,
\end{equation}
where $U_{n}(x)$ are Chebyshev polynomials of the second kind.

In Eq.~(\ref{eq56}), we can see that two eigenvalues are $u_{\pm 0}=\pm 1$ and the other eigenvalues are determined by $U_{{\scriptscriptstyle N}-2}(x)=0$.
The $N-2$ roots for $U_{{\scriptscriptstyle N}-2}(x)=0$ are $x_k=\cos\theta_k, \theta_k=k\pi/(N-1), k=1,2,\cdots,N-2$. Using the mapping relation $u=bx\pm i\sqrt{1-b^2x^2}$, the $2(N-2)$ eigenvalues are given by,
\begin{equation}\label{eq57} \small
u_{\pm k}=bx_k\pm i\sqrt{1-b^2x_k^2}, x_k=\cos \frac{k\pi}{N-1}, k\in [1,N-2].
\end{equation}

Now we analyze the eigenstates $|\Psi^s\rangle$. The eigenstates corresponding to eigenvalues $u_{\pm 0}=\pm 1$ can be easily determined by Eq.~(\ref{eq55}) and (\ref{eq53}). When $u_{\pm 0}=\pm 1$, $x_{\pm 0}=\pm 1/b$ (see the mapping relation). Eq.~(\ref{eq55}) becomes $\alpha_{3,L}^s=\pm\frac{1-a}{b}\alpha_{2,L}^s$, thus Eq.~(\ref{eq53}) can be written as,
\begin{equation}\label{eq58}
\begin{aligned}
\alpha_{i,L}^s(x_{\pm 0})=&\big[\pm \frac{1-a}{b}U_{i-3}(\pm\frac{1}{b})-U_{i-4}(\pm\frac{1}{b})\big]\alpha_{2,L}^s \\
=& \big(\pm \frac{1-a}{b}\big)^{i-2}\alpha_{2,L}^s, \ \ i\in [2,N]
\end{aligned}
\end{equation}
According to Eq.~(\ref{eq50}), the right components $\alpha_{i,R}^s$ of the eigenvalue $\pm 1$'s eigenstates can be recasted as,
\begin{equation}\label{eq59} \small
\begin{aligned}
\alpha_{i,R}^s(x_{\pm 0})=&\frac{b}{a}\alpha_{i,L}^s(x_{\pm 0})-\frac{u}{a}\alpha_{i+1,L}^s(x_{\pm 0}) \\
=&\big[\frac{b}{a}(\pm \frac{1-a}{b})^{i-2}-(\frac{\pm 1}{a})(\pm\frac{1-a}{b})^{i-1}\big]\alpha_{2,L}^s \\
=& \pm \big(\pm \frac{1-a}{b}\big)^{i-1}\alpha_{2,L}^s, \ \ i\in [1,N-1]
\end{aligned}
\end{equation}

For the eigenvalues satisfy $U_{{\scriptscriptstyle N}-2}(x)=0$, we can also express the corresponding eigenstates in terms of $\alpha_{2,L}^s$. Using the mapping relation, Eq.~(\ref{eq55}) can be simplified as,
\begin{equation}\label{eq60}
\begin{aligned}
\alpha_{3,L}^s=&\frac{1-au^2}{bu}\alpha_{2,L}^s=\big[\frac{1}{b}(u+\frac{1}{u})-\frac{1+a}{b}u\big]\alpha_{2,L}^s \\
=& (2x-\frac{1+a}{b}u)\alpha_{2,L}^s
\end{aligned}
\end{equation}
Substituting the above equation into Eq.~(\ref{eq53}), we get,
\begin{equation}\label{eq61}
\begin{aligned}
\alpha_{i,L}^s(x)=&\big[(2x-\frac{1+a}{b}u)U_{i-3}(x)-U_{i-4}(x)\big]\alpha_{2,L}^s \\
=&\big[U_{i-2}(x)-\frac{1+a}{b}uU_{i-3}(x)\big]\alpha_{2,L}^s,  i\in [2,N]
\end{aligned}
\end{equation}
Substituting the above equation into Eq.~(\ref{eq50}), we obtain the expression for $\alpha_{i,R}^s$,
\begin{equation}\label{eq62}
\begin{aligned}
\alpha_{i,R}^s(x)=&\frac{b}{a}\alpha_{i,L}^s(x)-\frac{u}{a}\alpha_{i+1,L}^s(x) \\
=&\big\{\frac{b}{a}(U_{i-2}(x)-\frac{1+a}{b}uU_{i-3}(x))- \\
 &\frac{u}{a}(U_{i-1}(x)-\frac{1+a}{b}uU_{i-2}(x)) \big\}\alpha_{2,L}^s
\end{aligned}
\end{equation}
Using the mapping relation $u^2=2bux-1$, the above equation is simplified to be,
\begin{equation}\label{eq63} \small
\begin{aligned}
&\alpha_{i,R}^s(x)=\frac{b}{a}\alpha_{i,L}^s(x)-\frac{u}{a}\alpha_{i+1,L}^s(x) \\
&=\big\{\frac{b}{a}U_{i-2}(x)-\frac{1+a}{a}uU_{i-3}(x)- \frac{u}{a}U_{i-1}(x)+ \\
& \frac{2xu}{a}U_{i-2}(x)+2xuU_{i-2}(x) -\frac{(1+a)}{ab}U_{i-2}(x) \big\}\alpha_{2,L}^s \\
&=\big\{(\frac{b}{a}-\frac{(1+a)}{ab})U_{i-2}(x))-\frac{u}{a}[U_{i-1}(x)+U_{i-3}(x)] \\
&+\frac{2xu}{a}U_{i-2}(x)+u[2xU_{i-2}(x)-U_{i-3}(x)]\big\}\alpha_{2,L}^s \\
&=\big\{-\frac{1+a}{b}U_{i-2}(x)+uU_{i-1}(x)\big\}\alpha_{2,L}^s, i\in [1,N-1]
\end{aligned}
\end{equation}
where the terms in square brackets are simplified using identify (\ref{a7}) in Appendix.

We have obtained all the eigenstates as a function of $\alpha_{2,L}^s$. Using the Chebyshev identities in Appendix A, one can prove the symmetric relation
$|\alpha_{i,L}^s|=|\alpha_{i-1,R}^s|$ ($i\in [2,N]$). It is worth mentioning that the symmetric property of the eigenvector components is different from the case of the moving shift operator (See Eqs.~(\ref{eq40})-(\ref{eq42})). To determine $|\alpha_{2,L}^s|$, we use the the normalization condition $\sum_{i,J}|\alpha_{i,J}^s|^2=\sum_{i=2}^{N}|\alpha_{i,L}^s|^2+\sum_{i=1}^{N-1}|\alpha_{i,R}^s|^2=1$. For the eigenvalues $u_{\pm0}=\pm 1$, we use Eqs.~(\ref{eq58}) and (\ref{eq59}) to calculate $|\alpha_{2,L}^s|$, which leads to,
\begin{equation}\label{eq64}
\begin{aligned}
|\alpha_{2,L}^s(x_{\pm 0})|^2=&(2\sum_{j=2}^N|\alpha_{i,L}^s(x_{\pm 0})|^2)^{-1} \\
=&\frac{1-r}{2(1-r^{N-1})}, \ \ \ r\equiv(\frac{1-a}{b})^2
\end{aligned}
\end{equation}
For eigenvalues in Eq.~(\ref{eq57}), $|\alpha_{2,L}^s(x_k)|$ can also be determined by the normalization condition $2\sum_{j=2}^N|\alpha_{i,L}^s(x_{k})|^2=1$.
Utilizing Eqs.~(\ref{eq61}) and (\ref{eq63}), after some algebraic calculus, we arrive at the following expression for $|\alpha_{2,L}^s(x_k)|$,
\begin{equation}\label{eq65}
\begin{aligned}
|\alpha_{2,L}^s(x_{\pm k})|^2=&\frac{\frac{1-a}{2}\sin^2\theta_k}{(N-1)(1-b^2\cos^2\theta_k)}, \\
&\theta_k=\cos\frac{k\pi}{N-1},k\in[1,N-2]
\end{aligned}
\end{equation}

\subsection{Long-time averaged probability distribution}
In this section, we use the eigenstates in the above to calculate the long-time averaged probability distribution. According to Eq.~(\ref{eq7}), the distribution can be written as,
\begin{equation}\label{eq66}
\begin{aligned}
\chi_{i,j}&= \sum_{k= \pm 0}^{\pm(N-2)}|\langle j, L|\Psi^s(x_k)\rangle|^2\cdot |\langle \Psi^s(x_k)|\psi_{i,0}\rangle|^2 \\
          & + \sum_{k=\pm 0}^{\pm(N-2)}|\langle j, R|\Psi^s(x_k)\rangle|^2\cdot |\langle \Psi^s(x_k)|\psi_{i,0}\rangle|^2 \\
          &= \sum_{k= \pm 0}(|\langle j, L|\Psi^s(x_k)\rangle|^2+|\langle j, R|\Psi^s(x_k)\rangle|^2) \cdot |\langle \Psi^s(x_k)|\psi_{i,0}\rangle|^2 + \\
          & \sum_{k=\pm 1}^{\pm(N-2)}(|\langle j, L|\Psi^s(x_k)\rangle|^2+|\langle j, R|\Psi^s(x_k)\rangle|^2)\cdot |\langle \Psi^s(x_k)|\psi_{i,0}\rangle|^2 \\
          &=2[|\alpha_{j,L}^s(x_{+0})|^2+|\alpha_{j,R}^s(x_{+0})|^2] \cdot |\langle \Psi^s(x_{+0})|\psi_{i,0}\rangle|^2 \\
          &+2\sum_{k=1}^{N-2}[|\alpha_{j,L}^s(x_{k})|^2+|\alpha_{j,R}^s(x_{k})|^2]\cdot |\langle \Psi^s(x_k)|\psi_{i,0}\rangle|^2 .
\end{aligned}
\end{equation}
In the above equation, the first term is the contribution from the eigenvalues $\pm 1$, and the second term is the contribution from the other eigenvalues. In the following, we will discuss $\chi_{i,j}$ according to the staring position and initial state $|\psi_{i,0}\rangle$ in Eq.~(\ref{eq12}). For the sake of simplicity, we first focus on the case $i=1$ (starting at the left most node), then we extend the conclusion to the general case. When the walk starts at $i=1$, the initial state $|\psi_{i,0}\rangle=|1,R\rangle$ (see Eq.~(\ref{eq12})), the probability of finding the particle at the original site $j=i=1$ only depends on the second term in the square brackets of Eq.~(\ref{eq66}),
\begin{equation}\label{eq67}
\begin{aligned}
\chi_{1,1}=&2|\alpha_{1,R}^s(x_{+0})|^4+2\sum_{k=1}^{N-2}|\alpha_{1,R}^s(x_{k})|^4\\
=&2|\alpha_{2,L}^s(x_{+0})|^4+2\sum_{k=1}^{N-2}|\alpha_{2,L}^s(x_{k})|^4 \\
=&2[\frac{1-r}{2(1-r^{N-1})}]^2+2\sum_{k=1}^{N-2} [\frac{\frac{1-a}{2}\sin^2\theta_k}{(N-1)(1-b^2\cos^2\theta_k)}]^2 \\
\approx &2[\frac{1-r}{2(1-r^{N-1})}]^2+\frac{(a+2)(1-a)^2}{4(a+1)^2}\cdot \frac{1}{N-1} \\
\approx &\frac{1}{2}\big(\frac{1-r}{1-r^{N-1}}\big)^2+{\cal O}(\frac{1}{N})
\end{aligned}
\end{equation}
The probability of finding the particle at the right most site $j=N$ only depends on the first term in the square brackets of Eq.~(\ref{eq66}),
\begin{equation}\label{eq68}
\begin{aligned}
\chi_{1,N}=&2|\alpha_{N,L}^s(x_{+0})|^2\cdot |\alpha_{1,R}^s(x_{+0})|^2 \\
  &+ 2\sum_{k=1}^{N-2}|\alpha_{N,L}^s(x_{k})|^2\cdot |\alpha_{1,R}^s(x_{k})|^2\\
=&2r^{N-2}|\alpha_{2,L}^s(x_{+0})|^4 \\
 &+2\sum_{k=1}^{N-2}|\frac{1+a}{b}|^2\cdot |U_{N-3}(x_k)|^2\cdot |\alpha_{2,L}^s(x_{k})|^4
\end{aligned}
\end{equation}
Noting that $U_{N-2}(x_k)=0$ ($k\in [1,N-2]$), $U_{N-3}(x_k)=U_{1}(x_k)=1$, the above equation transforms into,
\begin{equation}\label{eq69}
\begin{aligned}
\chi_{1,N}&=2r^{N-2}|\alpha_{2,L}^s(x_{+0})|^4 +2\sum_{k=1}^{N-2}|\frac{1+a}{b}|^2\cdot |\alpha_{2,L}^s(x_{k})|^4 \\
=&2r^{N-2}[\frac{1-r}{2(1-r^{N-1})}]^2+\frac{b^2/2}{(N-1)^2}\sum_{k=1}^{N-2}\frac{\sin^4\theta_k}{(1-b^2\cos^2\theta_k)^2} \\
\approx &\frac{1}{2}\big(\frac{1-r}{1-r^{N-1}}\big)^2r^{N-2}+\frac{(a+2)(1-a)}{4(1+a)}\cdot \frac{1}{N-1} \\
\approx &\frac{1}{2}\big(\frac{1-r}{1-r^{N-1}}\big)^2r^{N-2}+{\cal O}(\frac{1}{N})
\end{aligned}
\end{equation}
Similarly, the probability of finding the particle at the middle sites $j\in[2,N-1]$ is related to both the two terms in Eq.~(\ref{eq66}),
\begin{equation}\label{eq70}
\begin{aligned}
\chi_{1,j}\approx &\frac{1}{2}\big(\frac{1-r}{1-r^{N-1}}\big)^2(r^{j-2}+r^{j-1})+{\cal O}(\frac{1}{N}).
\end{aligned}
\end{equation}
Eqs.~(\ref{eq67}), (\ref{eq69}) and (\ref{eq70}) suggest that the long-time averaged probabilities are determined by two terms. The first term is determined by the eigenstates of the eigenvalues $\pm 1$, the second term is determined by the other continuous eigenstates and is inversely proportion to the system size $N$. When $N\rightarrow \infty$, the second term vanishes and the probability distribution only depends on the eigenstates of the eigenvalues $\pm1$. In this case, the eigenvalues $\pm 1$ play an important role in the probability distributions. This is different from the case of the moving shift operator.

When the system size is large, the probability distribution is mainly determined by the first term of Eq.~(\ref{eq66}). To achieve this, we calculate $|\alpha_{j,L}^s(x_{+0})|^2+|\alpha_{j,R}^s(x_{+0})|^2$ and $|\langle \Psi^s(x_{+0})|\psi_{i,0}\rangle|^2$ respectively. $|\alpha_{j,L}^s(x_{+0})|^2+|\alpha_{j,R}^s(x_{+0})|^2$ can be summarized as the following form,
\begin{equation}\label{eq71}
\begin{aligned}
f(r,j,N)=&|\alpha_{j,L}^s(x_{+0})|^2+|\alpha_{j,R}^s(x_{+0})|^2 \\
=&[(1-\delta_{1j})r^{j-2}+(1-\delta_{Nj})r^{j-1}]\cdot |\alpha_{2,L}^s(x_{+0})|^2
%=&[(1-\delta_{1j})r^{j-2}+(1-\delta_{Nj})r^{j-1}]\frac{1-r}{2(1-r^{N-1})}
\end{aligned}
\end{equation}

Supposing the initial state $|\psi_{i,0}\rangle$ takes the form in Eq.~(\ref{eq12}), the term $|\langle \Psi^s(x_{+0})|\psi_{i,0}\rangle|^2$ can be simplified as,
\begin{equation}\label{eq72}
\begin{aligned}
I(r,i,N)=&|\langle \Psi^s(x_{+0})|\psi_{i,0}\rangle|^2 \\
=&[(1-\delta_{1i}-\delta_{Ni})(r^{i-2}\cos^2\omega_0 + r^{i-1}\sin^2\omega_0 )\\
 & +\delta_{1i}r^{i-1}+\delta_{Ni}r^{i-2}]|\alpha_{2,L}^s(x_{+0})|^2
\end{aligned}
\end{equation}
Combining Eqs.~(\ref{eq71}) and (\ref{eq72}), we obtain a general expression for the probability $\chi_{i,j}$ in Eq.~(\ref{eq66}),
\begin{equation}\label{eq73}
\begin{aligned}
\chi_{i,j}=&2f(r,j,N)I(r,i,N)+{\cal O}(\frac{1}{N}) \\
=&\frac{1}{2}[(1-\delta_{1j})r^{j-2}+(1-\delta_{Nj})r^{j-1}] \\
 & \cdot [(1-\delta_{1i}-\delta_{Ni})(r^{i-2}\cos^2\omega_0 + r^{i-1}\sin^2\omega_0 )\\
 & +\delta_{1i}r^{i-1}+\delta_{Ni}r^{i-2}] (\frac{1-r}{1-r^{N-1}})^2+{\cal O}(\frac{1}{N})
\end{aligned}
\end{equation}

Now we discuss the characteristics of the probability distribution. First, we find that $\chi_{i,j}$ does not show symmetric feature. This differs from the moving shift operator where $\chi_{i,j}\equiv \chi_{i,N+1-j}$. For the swapping shift operator, such symmetry does not exist. Second, the first term of Eq.~(\ref{eq73}) converges to a nontrivial power-law stationary distribution when $r<1$ and $N\rightarrow \infty$. Here, we define the first term of Eq.~(\ref{eq73}) as the intrinsic probability, $\chi_{i,j}^{\rm Intri}=2f(r,j,N)I(r,i,N)$. The intrinsic probability only depends on the eigenstates of the eigenvalues $\pm 1$. For the case $r<1$ and $N\rightarrow \infty$, the total intrinsic probability $P_{\rm total}(i)=\sum_{j}\chi_{i,j}^{\rm Intri}$ can be written as,
\begin{equation}\label{eq74}
P_{\rm total}(i)=
\begin{cases}
1-r,  & \text{if } i=1 \\
(1-r)(r^{i-2}\cos^2\omega_0 + r^{i-1}\sin^2\omega_0 ),  & \text{if } i>1
\end{cases}
\end{equation}
This is one of the main conclusions in this paper. For the swapping shift operator, the probability distribution is not symmetric, the probability
distribution approaches to a power-law stationary distribution as as $r<1$ and $N\rightarrow \infty$. We show that such power-law stationary distribution is determined by the eigenstates of eigenvalues $\pm 1$ and discuss the condition for which the power-law stationary distribution occurs. The total intrinsic probability $P_{\rm total}(i)$ converges to a constant value in Eq.~(\ref{eq74}).

\section{CONCLUSIONS AND DISCUSSIONS}
In summary, we consider the discrete-time quantum walks on 1D Chain with the moving and swapping shift operators, respectively. We derive analytically expressions for the eigenvalues and eigenstates of the evolution operator using the Chebyshev polynomial technique, and calculate the
long-time averaged probabilities for the two different shift operators. It is found that the probability distributions for the moving and swapping shift operators display completely different characteristics. For the moving shift operator, the probability distribution exhibits high symmetry where the probabilities at mirror positions are equal. The probabilities are inversely proportional to the system size $N$ and approach to zero as $N\rightarrow \infty$. On the contrary, for the swapping shift operator, the probability distribution is not symmetric, the probability distribution approaches to a power-law stationary distribution as as $N\rightarrow \infty$ under certain condition of the coin parameter. We show that such
power-law stationary distribution is determined by the eigenstates of the eigenvalues $\pm 1$ and calculate the intrinsic probability $P_{\rm total}(i)$ for different starting positions. Our findings suggest that the eigenstates corresponding to eigenvalues $\pm 1$ play an important role for the dynamics of the swapping shift operator.

It is worth mentioning that the different dynamics of DTQWs on 1D chain using the moving and swapping shift operators are caused by the eigenvalues of the evolution operator $\hat{U}$. There are two special isolated eigenvalues $\pm 1$ for the swapping shift operator, while no special eigenvalues exist for the moving shift operator. This feature is similar to some dynamic processes taking place in networks where the largest or smallest eigenvalues play an important role in the relevant dynamics~\cite{rn20,rn21}. The eigenvalues $\pm 1$ determine the intrinsic probability $\chi_{i,j}^{\rm Intri}$, which shows a power-law behavior $\chi_{i,j}^{\rm Intri}\sim r^{j}$ ($j>1$) under the condition $r\equiv(1-a)^2/b^2<1$. For $r\geqslant 1$ and $N\rightarrow \infty$, the intrinsic probability vanishes. In Eq.~(\ref{eq9}), we use $|\phi\rangle=\cos\frac{1}{2}\omega|L\rangle+\sin\frac{1}{2}\omega|R\rangle$ to construct the coin operator $\hat{C}$. The condition $r\equiv(1-a)^2/b^2<1$ is equivalent to the condition $|\sin\frac{1}{2}\omega|^2<|\cos\frac{1}{2}\omega|^2$. This suggests that there is a nontrivial intrinsic probability when the left moving probability $|\cos\frac{1}{2}\omega|^2$ is larger than the right moving probability $|\sin\frac{1}{2}\omega|^2$. This could be a boundary effect of the left-most node of the 1D chain system in the spreading dynamics. When $|\sin\frac{1}{2}\omega|^2\geqslant|\cos\frac{1}{2}\omega|^2$, the particle are absorbed into the right-most positions and the intrinsic probability tends to 0. In our work, the Chebyshev polynomial technique is a good analytical tool to treat the problem, we believe the technique widely used in this paper will shed some light on the analytical calculations of the problems of quantum walks.

\begin{acknowledgments}
This work is supported by the National Natural Science Foundation of China under project 11205110, Korean National Research Foundation (NRF) Grant No. 20110029457, and Shanghai Key Laboratory of Intelligent Information Processing (IIPL-2011-009). Yusuke Ide is supported by the Grant-in-Aid for Young Scientists (B) of Japan Society for the Promotion of Science (Grant No. 23740093). Norio Konno is supported by the Grant-in-Aid for Scientific Research (C) of Japan Society for the Promotion of Science (Grant No. 24540116).
\end{acknowledgments}
\appendix
\onecolumngrid
\section{Definition of Chebyshev polynomials}
The Chebyshev polynomials of the first kind are defined by the recurrence relation~\cite{rn29,rn30},
\begin{equation}\label{a1}
T_0(x)=1, T_1(x)=x, 2xT_n(x)=T_{n-1}(x)+T_{n+1}(x).
\end{equation}
The Chebyshev polynomials of the second kind are defined by the recurrence relation~\cite{rn29,rn30},
\begin{equation}\label{a2}
U_0(x)=1, U_1(x)=2x, 2xU_n(x)=U_{n-1}(x)+U_{n+1}(x).
\end{equation}
The closed-form solutions of Eqs.~(\ref{a1}) and (\ref{a2}) are given by,
\begin{align}
T_n(x)&=\frac{z^n+z^{-n}}{2}, \label{a3} \\
U_n(x)&=\frac{z^{-(n+1)}-z^{n+1}}{|z^{-1}-z|}, \label{a4}
\end{align}
where $z=x-\sqrt{x^2-1}$. For the case of large order $n$ and $z<-1$ ($|z|>1$), the above solutions can be approximated by,
\begin{equation}\label{a5}
T_n(x)\approx\frac{z^n}{2}, U_n(x)\approx-\frac{z^{n+1}}{|z^{-1}-z|}=-\frac{z^{n+1}}{z^{-1}-z}
\end{equation}
Using the closed-form solutions for the Chebyshev polynomials (See Eqs.~(\ref{a3}) and (\ref{a4})), we can prove the following identities,
\begin{eqnarray}
U_{n-1}(x)+U_{-n-1}(x)=0,  \label{a6} \\
2xU_n(x)=U_{n-1}(x)+U_{n+1}(x), \label{a7}\\
T_n(x)=U_n(x)-xU_{n-1}(x)=xU_{n-1}(x)-U_{n-2}(x), \label{a8} \\
U_n(x)U_m(x)-U_{n-1}(x)U_{m-1}(x)=U_{n+m}(x), \label{a9}   \\
U_n(x)U_m(x)-U_{n+1}(x)U_{m-1}(x)=U_{n-m}(x), \label{a10} \\
U_n(x)T_m(x)+U_{m-1}(x)T_{n+1}(x)=U_{n+m}(x), \label{a11}  \\
U_n(x)T_m(x)-U_{m-1}(x)T_{n+1}(x)=U_{n-m}(x), \label{a12} \\
T_n^2(x)-(x^2-1)U_{n-1}^2(x)=1,  \label{a13} \\
T_m(x)U_n(x)=\frac{1}{2}[U_{m+n}(x)+U_{n-m}(x)], \ \ T_m(x)T_n(x)=\frac{1}{2}[T_{m+n}(x)+T_{|m-n|}(x)]  . \label{a14}
\end{eqnarray}
\section{Definition of the Variant Chebyshev polynomials}
The variant Chebyshev polynomials are defined by the following recurrence relation,
\begin{equation}\label{b1}
V_{0}(y)=1, V_{1}(y)=2y, 2yV_{n}(y)=V_{n+1}(y)-V_{n-1}(y)
\end{equation}
The closed-form solution for the above equation can be related to the Chebyshev polynomials of the second kind,
\begin{equation}\label{b2}
V_{n}(y)=i^nU_n(x), y=ix
\end{equation}
\section{Calculation for the determinant equation Eq.~(\ref{eq33})}
For the sake of simplicity, we first simplify the four coefficients $c_1$, $c_2$, $c_3$ and $c_4$ in Eqs.~(\ref{eq30}) and (\ref{eq32}) using the mapping relation $u^2=2ayu+1=2iaxu+1$. The two coefficients in Eq.~(\ref{eq32}) can be simplified as,
\begin{equation}\label{c1}
c_1=au(u^2-b)=au(2iaxu+1-b)=ua(1-b)[1-4(1+b)x^2]+2ixa^2
\end{equation}
\begin{equation}\label{c2}
\begin{aligned}
c_2=&(u^2-b)(u^2-b^2)+a^2b=(2iaxu+1-b)(2iaxu+1-b^2)+a^2b \\
=& -4a^2x^2u^2+2iaxu(2-b-b^2)+(1-b)(1-b^2)+a^2b  \\
=&-4a^2x^2(2iaxu+1)+2iaxu(2-b-b^2)+a^2 \\
=&(1-b)\{2iuax[-4(b+1)x^2+b+2]+(1+b)(1-4x^2) \}
\end{aligned}
\end{equation}
Analogously, the two coefficients in Eq.~(\ref{eq30}) can be simplified as,
\begin{equation}\label{c3}
c_3=(1-bu^2)V_{N-4}(y)-auV_{N-5}(y)=i^{N-5}\{ u[2abxU_{N-4}(x)-aU_{N-5}(x)]+ i(1-b)U_{N-4}(x)  \}
\end{equation}
\begin{equation}\label{c4}
c_4=auV_{N-6}(y)-(1-bu^2)V_{N-5}(y)=i^{N-5}\{ ui[2abxU_{N-5}(x)-aU_{N-6}(x)]-(1-b)U_{N-5}(x)  \}
\end{equation}
To calculate the determinant equation $c_2c_3-c_1c_4$, we need to expand the products $c_2c_3$ and $c_1c_4$. For the $u^2$ term, we use the mapping relation $u^2=2iaxu+1$ to reduce the power. We also use the Chebyshev polynomial identity (\ref{a7}) $U_{N-4}(x)=2xU_{N-5}(x)-U_{N-6}(x)$ to replace the term $U_{N-4}(x)$ and replace $a^2$ with $a^2=1-b^2$ in the expanded result. After some cumbersome calculations,  $c_2c_3-c_1c_4$ can be simplified as,
\begin{equation}\label{c5}
c_2c_3-c_1c_4=i^{N-5}2ab(1-b)[2axi+u(1-4a^2x^2)]\{ (8bx^4+8x^4-4bx^2-8x^2+1)U_{N-5}(x)-x(4bx^2+4x^2-b-3)U_{N-6}(x)  \}
\end{equation}
The determinant equation requires $c_2c_3-c_1c_4=0$. There is no solution when the term in square bracket equals to 0, thus the solutions are determined by the term in the brace $\{ \}$ equals to 0. Consequently, the solutions of Eq.~(\ref{c5}) are given by,
\begin{equation}\label{c6}
(8bx^4+8x^4-4bx^2-8x^2+1)U_{N-5}(x)-x(4bx^2+4x^2-b-3)U_{N-6}(x)=0
\end{equation}
The polynomials in the Chebyshev polynomials can be written as,
\begin{equation}\label{c7}
(8bx^4+8x^4-4bx^2-8x^2+1)=\frac{1}{2}(b+1)(16x^4-12x^2+1)+\frac{1}{2}(b-1)(4x^2-1)=\frac{1}{2}(b+1)U_4(x)+\frac{1}{2}(b-1)U_2(x)
\end{equation}
\begin{equation}\label{c8}
x(4bx^2+4x^2-b-3)=\frac{1}{2}(b+1)[x(8x^2-4)]+\frac{1}{2}(b-1)(2x)=\frac{1}{2}(b+1)U_3(x)+\frac{1}{2}(b-1)U_1(x)
\end{equation}
Substituting Eqs.~(\ref{c7}) and (\ref{c8}) into Eq.~(\ref{c6}), we have
\begin{equation}\label{c9}
\frac{1}{2}(b+1)[U_4(x)U_{N-5}(x)-U_3(x)U_{N-6}(x)]+\frac{1}{2}(b-1)[U_2(x)U_{N-5}(x)-U_1(x)U_{N-6}(x)]=0
\end{equation}
Using the Chebyshev polynomial identity (\ref{a9}), the terms in the square brackets are equal to $U_{N-1}(x)$ and $U_{N-3}(x)$ respectively. Hence, the determinant equation becomes,
\begin{equation}\label{c10}
U_{N-1}(x)+\frac{b-1}{b+1}U_{N-3}(x)=0
\end{equation}

\section{Calculation for the determinant equation Eq.~(\ref{eq56})}
In this section, we show the determinant of the Eqs.~(\ref{eq54}) and (\ref{eq55}) can be simplified into Eq.~(\ref{eq56}). The determinant of the coefficients can be written as,
\begin{equation}\label{d1}
bu[(a+u^2)U_{N-4}(x)-buU_{N-5}(x)]-(1-au^2)[(a+u^2)U_{N-3}(x)-buU_{N-4}(x)]=0
\end{equation}
Substituting $U_{N-5}(x)=2xU_{N-4}(x)-U_{N-3}(x)$ (See Identity (\ref{a7})) into the above equation leads to,
\begin{equation}\label{d2}
\begin{aligned}
&bu[(a+u^2)U_{N-4}(x)-bu(2xU_{N-4}(x)-U_{N-3}(x))]-(1-au^2)[(a+u^2)U_{N-3}(x)-buU_{N-4}(x)]\\
&=[ b^2u^2-(1-au^2)(a+u^2)]U_{N-3}(x)+bu[(a+u^2-2bux)+(1-au^2)]U_{N-4}(x) \\
&=[u^2(a^2+b^2)-u^2-a+au^4]U_{N-3}(x)+bu[(a+u^2-(u^2+1))+(1-au^2)]U_{N-4}(x) \\
&=a(u^2-1)(u^2+1)U_{N-3}(x)-abu(u^2-1)U_{N-4}(x) \\
&=abu(u^2-1)[2xU_{N-3}(x)-U_{N-4}(x)]=abu(u^2-1)U_{N-2}(x)=0
\end{aligned}
\end{equation}
where the mapping relation $2bxu=u^2+1$ and Identity (\ref{a7}) are used in the above calculation.
\twocolumngrid
%\begin{figure}
%\scalebox{0.28}[0.28]{\includegraphics{Graph6a}}
%\scalebox{0.28}[0.28]{\includegraphics{Graph6b}}
%\scalebox{0.28}[0.28]{\includegraphics{Graph6c}}%
%\caption{(Color online) The states of the communities. The circle
%and the square present two sort of community and the communities
%with the same color indicate they are synchronized. (a), (b) and (c)
%represent the cases of $C=0.01$, 0.1 and 0.3, as shown in the above
%numerical simulations.}
%\end{figure}

\end{document}